\documentclass[prl,onecolumn,preprintnumbers,superscriptaddress, bibnotes,nofootinbib]{revtex4}

\usepackage{amsthm}
\usepackage{amsmath}
\usepackage{amssymb}
\usepackage{graphicx}
\usepackage{dsfont}
\usepackage{float}
\usepackage{bm}
\usepackage{ifthen}
\usepackage[dvipsnames]{xcolor}
\usepackage{mathrsfs}
\usepackage[colorlinks=true,citecolor=blue,urlcolor=black]{hyperref}
\usepackage{float}
\usepackage{afterpage}
\usepackage{subfigure}
\usepackage{adjustbox}
\usepackage{times}
\usepackage{multirow}
\usepackage{tabularx}

\begin{document}

	\title{Proof-of-principle experimental demonstration of twin-field quantum key distribution over optical channels with asymmetric losses}
	
	\author{Xiaoqing Zhong}
	\affiliation{Center for Quantum Information and Quantum Control, Dept. of Physics, University of Toronto, Toronto, Ontario, M5S 1A7, Canada}
	
	\author{Wenyuan Wang}
	\affiliation{Center for Quantum Information and Quantum Control, Dept. of Physics, University of Toronto, Toronto, Ontario, M5S 1A7, Canada}
	
	\author{Li Qian}
	\affiliation{Center for Quantum Information and Quantum Control, Dept. of Electrical \& Computer Engineering, University of Toronto, Toronto, Ontario, M5S 3G4, Canada}
	
	\author{Hoi-Kwong Lo}
	\affiliation{Center for Quantum Information and Quantum Control, Dept. of Physics, University of Toronto, Toronto, Ontario, M5S 1A7, Canada}
	\affiliation{Center for Quantum Information and Quantum Control, Dept. of Electrical \& Computer Engineering, University of Toronto, Toronto, Ontario, M5S 3G4, Canada}		

\begin{abstract}
	Twin-field (TF) quantum key distribution (QKD) is highly attractive because it can beat the fundamental limit of secret key rate for point-to-point QKD without quantum repeaters. Many theoretical and experimental studies have shown the superiority of TFQKD in long-distance communication. All previous experimental implementations of TFQKD have been done over optical channels with symmetric losses. But in reality, especially in a network setting, the distances between users and the middle node could be very different. In this paper, we perform a first proof-of-principle experimental demonstration of TFQKD over optical channels with asymmetric losses. We compare two compensation strategies, that are (1) applying asymmetric signal intensities and (2) adding extra losses, and verify that strategy (1) provides much better key rate. Moreover, the higher the loss, the more key rate enhancement it can achieve. By applying asymmetric signal intensities, TFQKD with asymmetric channel losses not only surpasses the fundamental limit of key rate of point-to-point QKD for 50 dB overall loss, but also has key rate as high as $2.918\times10^{-6}$ for 56 dB overall loss. Whereas no keys are obtained with strategy (2) for 56 dB loss. The increased key rate and enlarged distance coverage of TFQKD with asymmetric channel losses guarantee its superiority in long-distance quantum networks. 
\end{abstract}
	\maketitle
			
	Quantum key distribution (QKD) enables remote users to share secret keys with information-theoretic security~\cite{R1,R2}. However, due to the unavoidable losses of optical channels, there exists a fundamental limit on the achievable secret key rate of long distance QKD. Without using quantum repeaters, the upper bound (also called repeaterless bound in this paper) of the secret key rate of QKD scales linearly with the channel transmittance $\eta$~\cite{plob,R21}. Remarkably, a new type of QKD, called twin-field (TF) QKD, has been proposed~\cite{tf-qkd_original} and can practically overcome the  repeaterless bound. In TFQKD, like in the measurement-device-independent (MDI) QKD~\cite{mdiqkd}, two users (Alice and Bob) send two coherent states to an un-trusted intermediate node, i.e. Charlie, who performs the measurement. Because TFQKD employs single photon interference, rather than two-photon interference in MDIQKD, the secret key rate of TFQKD scales as $\sqrt{\eta}$, allowing for unprecedented distance coverage. Plenty of variations and security analysis of TFQKD~\cite{R39,R40,R41,R42,tf-qkd_marcos,R44} have been studied, followed by multiple experimental demonstrations~\cite{minder,our_paper,e1,e2}. More recently, TFQKD has been successfully implemented over more than 500 km fibers~\cite{new1,new2}. It has been shown that TFQKD is one of the most promising and practical solutions to long distance QKD.   
	  
	 \begin{figure}[b]
		 	\includegraphics[width=0.65\linewidth]{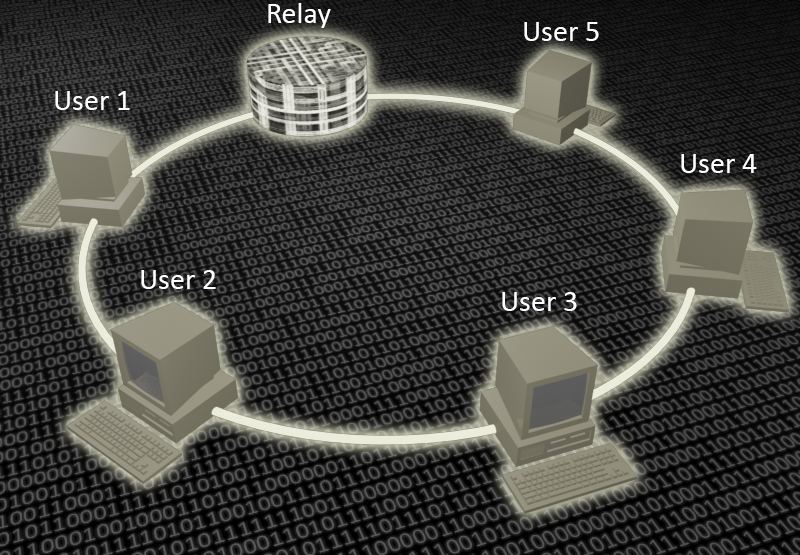}	
		 	\caption{Illustration of a Sagnac-loop-based TFQKD network. Multiple users can be placed on the same loop to communicate via a single relay. As can be seen here, arbitrary pairs of users can have very different distances (channel losses) from the relay, which necessitate a TFQKD protocol that maintains good performance even in the presence of channel asymmetry. In this work we present the experimental implementation of an asymmetric-intensity TFQKD protocol that maintains high rate through asymmetric channels, thus demonstrating the feasibility of such a Sagnac-loop-based TFQKD network.} 
		 	\label{network}
	 \end{figure}
 
	However, all the above mentioned studies only consider TFQKD over optical channels with symmetric losses between each of the users and intermediate node, and let Alice and Bob use identical sets of operations in preparing their signals. However, this assumption on channel symmetry is seldom true in reality. TFQKD over asymmetric channels is important not only for practical point-to-point implementations, but also in a network setting in where the optical distances between users and the middle node can be significantly different. For instance, as shown in Fig.(\ref{network}), if we consider a Sagnac-loop setup, multiple users can be placed on the same loop, where they share a common relay, to implement a TFQKD network. However, the users on the loop naturally will have different distances to the relay, thus making asymmetric channels a major characteristic for such a TFQKD network setup. Similar problems also exist for star-shaped networks where users are placed arbitrary distances away from a central relay.
	
	Unfortunately, because TFQKD depends on a good visibility of single-photon interference, it requires the two channels to have similar levels of loss. This means that current implementations of TFQKD will have suboptimal or even zero key rate if channels are asymmetric. One intuitive solution is to deliberately add fibers/losses to compensate for the shorter distance~\cite{add}. But this solution is not the optimal strategy, because it would increase signal loss and thus lower the secret key rate. 
	
	Instead of physically adding fibres/losses, several recent theoretical papers \cite{asy-tfqkd1,asy-tfqkd2,asy-tfqkd3,asy-tfqkd4} in TFQKD study the use of asymmetric intensities between Alice and Bob to compensate for channel asymmetry and obtain optical secret key rate\footnote{The limitation to symmetric optical channels has been first observed and investigated for MDIQKD, whose visibility also requires symmetry between optical channels. Refs.\cite{practical_mdiqkd,asy-mdiqkd1} proposed a method to compensate for channel asymmetry in MDIQKD with asymmetric laser intensities, which is experimentally verified in Ref.\cite{asy-mdiqkd2}.}. Ref.\cite{asy-tfqkd1,asy-tfqkd4} are based on an asymmetric-intensity version of the "Sending-or-not-Sending (SNS)" Protocol \cite{R41}, while Refs.\cite{asy-tfqkd2,asy-tfqkd3} are based on the protocol proposed in \cite{tf-qkd_marcos} by Curty, Azuma, Lo (for simplicity, let us call the protocol "CAL19" protocol here). 
		
	In this paper we have implemented the protocol in Ref.\cite{asy-tfqkd3}\footnote {We choose Ref. \cite{asy-tfqkd3} over the other three references, as the CAL19 protocol provides higher key rate than SNS protocol except over extremely long distances. Also, while Ref.\cite{asy-tfqkd2,asy-tfqkd3} are both based on the CAL19 protocol, Ref.\cite{asy-tfqkd3} provides the additional convenience of only requiring signal states (and not decoy states) to be asymmetric.}. The key point of the protocol is that Alice and Bob can adjust their signal intensities independently, to effectively compensate for channel asymmetry. In this work, we for the first time experimentally demonstrate TFQKD over optical channels with asymmetric losses, and show that the new protocol provides much higher key rate and longer distance than previous strategies (adding extra loss or using no compensation at all). Importantly, this also shows the feasibility of a TFQKD based quantum network. 
	
	The key steps of the asymmetric-intensity TFQKD protocol\cite{asy-tfqkd3} demonstrated in this paper are summarized as follows. Alice and Bob prepare weak coherent states and randomly choose X and Z bases. For signal states in X basis, Alice and Bob randomly add a $0$ or $\pi$ phase and set the intensities of states to $s_A$ and $s_B$ respectively. For decoy states in Z basis, a random phase is added and the intensities are randomly chosen from $\left\lbrace \mu,\nu,\omega\right\rbrace $. The important difference from the CAL19 protocol in Ref.\cite{tf-qkd_marcos,our_paper} is that, signal intensities $s_A$ and $s_B$ can be set to different values, while the intensities of Alice's and Bob's decoy states are still kept symmetric. Such a choice of intensities is because, as explained in Ref.\cite{asy-tfqkd3}, the X basis requires intensities arriving at Charles to be symmetric for a good interference visibility (hence $s_A,s_B$ should be different, to compensate for channel asymmetry), while the Z basis doesn't have such a requirement (hence decoy states can simply be set to symmetric to simplify implementation). The latter is because the estimation of phase error rate is based on photon-number yields in the Z basis, which is little affected by asymmetry of intensities arriving at Charles. Then Alice and Bob send their states to the middle node Charlie, who measures the interference of the coming states and announces the results. 
	  
\begin{figure}[b]
		\includegraphics[width=0.65\linewidth]{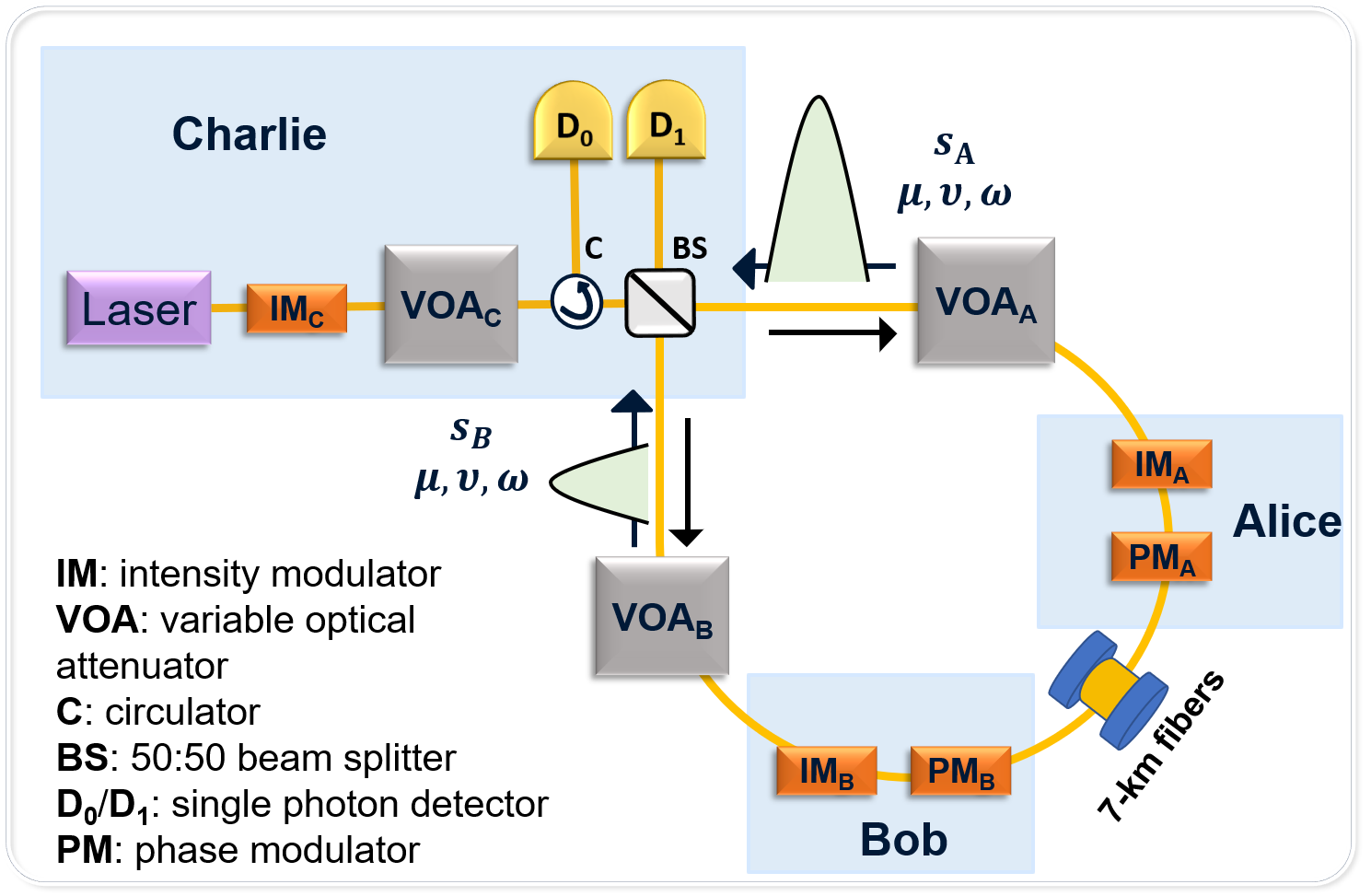}	
		\caption{Schematic set-up of twin-field quantum key distribution (TFQKD) over optical channels with asymmetric losses.  Charlie produces un-modulated weak coherent pulses through his intensity modulator (IM$_C$) and variable optical attenuator (VOA$_C$) and distributes the pulses to Alice and Bob. The pulses enter the Sagnac loop through a circulator and a 50:50 beam splitter (BS) where they splits into clockwise and counter-clockwise traveling pulses. The clockwise traveling pulses first pass through the attenuator VOA$_A$ and and Alice's station without being modulated. Then they go through a 7-km fiber spool and arrive at Bob's station.  Based on the bases Bob chooses (signal basis or decoy basis), he modulates the phases and intensities of the pulses by his phase modulator and intensity modulator. Then Bob forwards the modulated pulses  back to Charlie's BS though the attenuator VOA$_B$. The same process applies to the counter-clockwise traveling pluses, except that only Alice would modulate the phases and intensities of the counter-clockwise traveling pulses. The modulated pulses from Alice and Bob interfere with each other at Charlie's BS and are detected by Charlie's two single photon detectors $D_0$ and $D_1$.} 
		\label{setup}
	\end{figure}

	The experimental setup used in our previous work of TFQKD over symmetric optical channels in Ref.\cite{our_paper} can be simply adopted to implement the asymmetric-intensity TFQKD protocol, as shown in Fig.(\ref{setup}). As a proof-of-principle demonstration, we only use the optical variable attenuators (VOA$_A$ and VOA$_B$) to simulate the optical channel losses between Alice/Bob and Charlie. The only difference from the setup in Ref.\cite{our_paper} is that VOA$_A$ and VOA$_B$ have different attenuation to mimic the  asymmetric channel losses. As indicated in Ref.\cite{our_paper}, a two-way QKD system consisting of a Sagnac interferometer is applied to overcome the main challenge of implementing TFQKD, namely, the phase stabilization. The common-path nature of the Sagnac loop automatically compensates for phase fluctuations, thus maintaining long-term phase stability between the weak coherent states sent from Alice and Bob. Moreover, the laser located on Charlie's station is shared by Alice and Bob, to guarantee the matched global phase. Filters, attenuators and monitoring detectors can be added on Alice's and Bob's stations to prevent possible attacks from eavesdropper. More discussion about the security of the setup can be found in Ref.\cite{our_paper}.
	
	Charlie uses his intensity modulator (IM$_C$) and VOA$_C$ to create weak coherent pulses (10MHz, 900 ps) from a continuous wave source and sends the pulses to Alice and Bob. The pulses go through an optical circulator and enter the Sagnac loop through a 50:50 beam splitter (BS), where the pulses splits into  clockwise traveling and counter-clockwise traveling pluses. Clockwise (counter-clockwise) pluses first go through VOA$_A$ (VOA$_B$) and Alice’s (Bob's) station without being modulated. Then the clockwise (counter-clockwise) pulses pass a 7-km fiber spool (with loss of about 7 dB) before reaching Bob's (Alice's) station. Note that no information is transmitted over the channel Between Alice and Bob. On Bob's (Alice's) station, the pulses are modulated by a phase modulator PM$_B$ (PM$_A$) and an intensity modulator IM$_B$ (IM$_A$). Based on different bases Alice and Bob choose, the phases and intensities of the pulses are modulated accordingly. All the modulators in the set-up are driven and synchronized by a high speed arbitrary waveform generator (AWG, Keysight M8195). The modulated pulses from Alice and Bob travel through the attenuators VOA$_B$ and VOA$_A$ and interfere at Charlie's BS.  One output of the BS is directed to a single-photon detector (SPD) $D_0$ via the circulator, and the other output is followed directly by another SPD, $D_1$. Charlie then uses $D_0$ and $D_1$ to record the interference and publicly announces the results. The SPDs used in the set-up are the commercial free-run avalanche photodiodes (ID220)), the dark count probability of which is about $7\times10^{-7}$. 
	
    It is very important to ensure that Alice and Bob only modulate the pulses traveling in designed directions. That is to say, the clockwise and counter-clockwise traveling pulses should never overlap with each other at any of Alice's and Bob's modulators. Therefore, the fiber lengths among the users and middle node are carefully calibrated to avoid the overlap of the arriving time at any modulators between the clockwise and counter-clockwise traveling pulses. Another challenge in experiment is that the limited extinction ratio of a single intensity modulator is not sufficient to generate the vacuum state ($\omega$), especially on Alice's station where the power of the injected pulse (that should be modulated) is always 10 dB higher than that on Bob's station. To create the vacuum state, we use two intensity modulators to achieve more than 65dB extinction ratio. The resulting pulse is suppressed below the dark count noise of the detectors. Multiple polarization controllers are used for the initial polarization alignment but no active polarization control is needed. Because of the auto compensation of phase fluctuation of Sagnac interferometer, our system is stable and the interference visibility is kept as high as 99.8\%. In this paper, the main objective is to study the optimal compensation strategy for TFQKD over asymmetric channels. Therefore, variable optical attenuators are used instead of real fibers. Since the ability of Sagnac loop withstanding phase fluctuations is a function of its total length and the characteristic frequencies of the fluctuations, when hundred of kilometers of real fibers are inserted into the loop to replace VOAs, the phase stability and polarization stability of the current system would be affected. However, previous study in Ref.\cite{our_paper} have found that a Sagnac loop with 300 km loop length, corresponding to 60 dB of loss, is adequate in maintaining phase stability required for TFQKD.
    
    \begin{table*}[t]
    	\begin{tabular}{@{\extracolsep{8pt}}cc|ccc|ccc|cc@{}}
    		\hline
    		\hline
    		{\bf Overall Loss} & & \multicolumn{6}{c|}{\bf Intensity } & \multicolumn{2}{c}{\bf Key rate}  \\
    		\hline
    		
    		{ }& & $s_A$ & $\mu_A$ & $\nu_A$ & $s_B$ & $\mu_B$ & $\nu_B$ & \bf{Infinite data}& \bf{Finite data}\\
    		\hline
    		
    		\multirow{3}{*}{ $25+15$ dB } & Asym. & $0.0448$ & $0.300$ & $0.120$ & $0.00529$ & $0.300$ & $0.120$ & $1.017\times10^{-4}$ & $5.013\times10^{-5}$\\
    		& Add loss & $0.0213$ & $0.481$ & $0.146$ &  $0.0213$ & $0.481$ & $0.146$ & $3.727\times10^{-5}$ & $1.688\times10^{-5}$ \\
    		& No comp. & $0.0036$ & $0.247$ & $0.0923$ & $0.0036$ & $0.247$ & $0.0923$ &$7.163\times10^{-6}$ & 0 \\
    		
    		\hline
    		\multirow{2}{*}{ $30+20$ dB } & Asym. & $0.030$ & $0.514$ & $0.108$ & $0.00373$ & $0.514$ & $0.108$ & $1.666\times10^{-5}$ & $6.971\times10^{-6}$\\
    		& Add loss & $0.0147$ & $0.444$ & $0.133$ & $0.0147$ & $0.444$ & $0.133$ & $2.382\times10^{-6}$& $2.677\times10^{-7}$ \\
    		\hline
    		
    		$33+23$ dB & Asym. & $0.0274$ & $0.401$ & $0.120$ & $0.0035$ & $0.401$ & $0.120$ & $2.918\times10^{-6}$ & $3.174\times10^{-7}$\\

    		\hline
    		\hline
    	\end{tabular}
    	\caption{\label{tab1} List of intensity sets and experimental secret key rates for the overall system loss 40 dB, 50 dB and 56 dB. The loss between Alice and Charlie is always 10 dB higher than the loss between Bob and Charlie. $s_{A/B}$ is Alice's/Bob's signal intensity. $\mu_{A/B}, \nu_{A/B}$ are the decoy intensities. $\omega$ is the vacuum state and is not listed here. The secret key rate is calculated based on the observed gains and quantum bit error rates. The size of the total data sent to Charlie is $3\times10^{10}$. Both infinite-data case and finite-data case are considered. For each loss, the first row shows intensities and key rates with asymmetric signal intensities; the second row (if exists) gives the intensities and key rates with adding extra losses; the third row (if exists) is the case where no compensation is applied.}
    \end{table*}
    
    \begin{figure*}
    	\includegraphics[width=\linewidth]{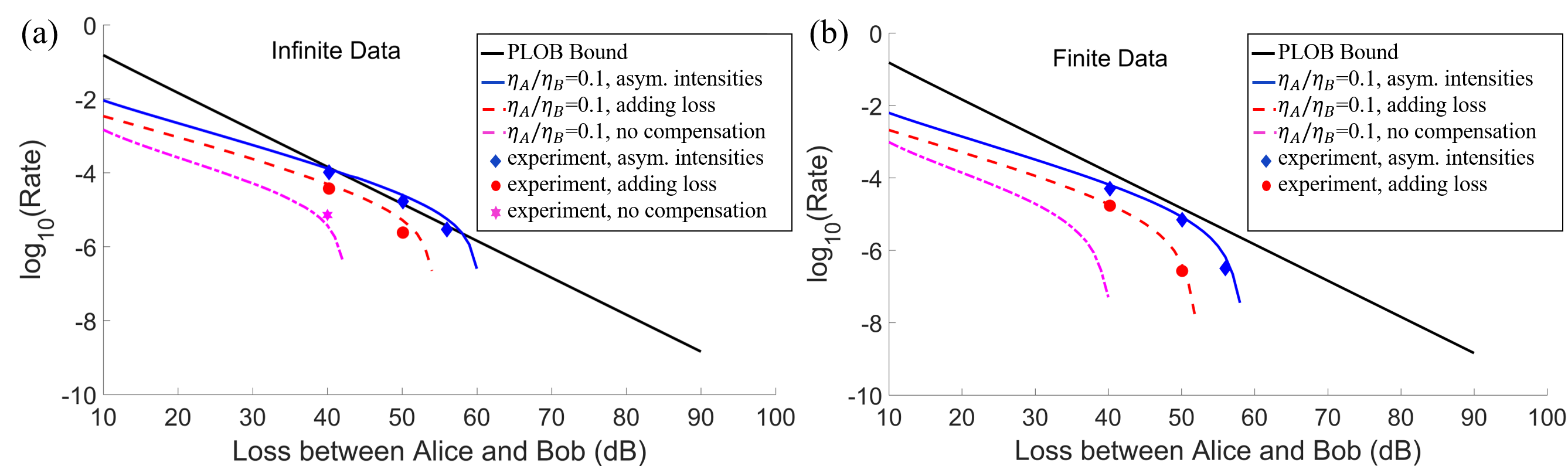}	
    	\caption{Secret key rate (per pulse) in logarithmic scale as a function of the overall loss between Alice and Bob for \textbf{a)} infinite-data case and \textbf{b)} finite-data case. The secret key rate is calculated based on the observed gains and quantum bit error rates. The size of the total data sent to Charlie is $3\times10^{10}$. The solid black line represents one representative of the repeaterless bound~\cite{plob}. The blue solid curve is the simulated key rate with asymmetric signal intensities; the red dash curve is the simulated key rate with adding extra losses; the purple dash-dotted curve is the simulated key rate with no compensation. All the scattered points are the experimental secret key rates. The blue rhombi represent the case with asymmetric signal intensities; the red circles represent the case where extra 10 dB attenuation is added on Bob's side; the purple hexagon is the key rate obtained when no compensation is applied. As observer, the strategy of applying asymmetric signal intensities always provides better key rates than the other two strategies.} 
    	\label{data}
    \end{figure*}
    
	The experiment has been performed over three overall channel losses between Alice and Bob, 40 dB, 50 dB and 56 dB. The channel losses between Alice and charlie is always 10 dB higher than the losses between Bob and Charlie, i.e., $\eta_B=\eta_A\times10$. The detector efficiency (11.7\%) is included in the overall loss. To test the asymmetric-intensity strategy, We allow Alice and Bob to choose asymmetric signal intensities $s_A$ and $s_B$, but keep their decoy intensities symmetric. We have also tested the strategy where all the intensities are symmetric but another 10 dB attenuation is added on Bob's side to compensate the channel asymmetry. Additionally, at the overall loss of 40 dB, we have conducted the experiment where Alice and Bob use identical sets of operations as they do for TFQKD with symmetric channels (no compensation at all).  All the signal intensities $s_{A/B}$ and decoy intensities {$\mu_{A/B}, \nu_{A/B}$} used in the experiment are close to the optimal values and are listed in table~(\ref{tab1}). ($\omega$ is the vacuum state and therefore is not listed.) Note that when Alice and Bob test the asymmetric-intensity strategy, the intuitive way is to set $s_A/s_B=\eta_B/\eta_A=10$. However, as indicated in table~(\ref{tab1}), the ratio of the optimal $s_A$ to $s_B$ slightly deviates from $10$. This is because, as described in Ref.\cite{asy-tfqkd3}, although the interference visibility (which affects X basis QBER) favors $s_A/s_B=\eta_B/\eta_A$, there are other factors affecting $s_A,s_B$ - namely, a tight estimation of the phase error rate favors small values of both $s_A$ and $s_B$ (which determine the cat state coefficients) and makes optimal $s_A/s_B$ deviate from exactly $\eta_B/\eta_A$.\footnote{We would like to point out that it is more convenient to use the intensities that fulfill $s_A/s_B=\eta_B/\eta_A$, especially for the Sagnac-loop based system which automatically provides such intensity compensation. Considering the experimental fluctuations, the tested key rate with the exact ratio $s_A/s_B=10$ can be even higher than the rate with optimal ratio.} The size of the total data that Alice and Bob send to Charlie in each run is $3\times10^{10}$. Due to the limit of the available AWG channels, the signal state and decoy state are not randomly switched in the experiment. But this random switch can be easily accomplished with more resources. As a proof-of-principle demonstration, our current implementation is feasible to study the optimal compensation strategies for TFQKD with asymmetric channel losses. The secret key rate is calculated based on the observed gains and quantum bit error rates. Both infinite-data case and finite-data case are considered and the experimental results are depicted in Fig.~(\ref{data}), which shows the secret key rate (per pulse) in logarithmic scale as a function of the overall loss between Alice and Bob. The blue rhombi are the experimental key rates obtained with asymmetric signal intensities; the red circles are the key rates of the case where extra 10 dB attenuation is added on Bob's side; the purple hexagon is the key rate obtained when no compensation is applied. The corresponding simulated secret key rates of the above three cases are also shown in Fig.~(\ref{data}), represented by blue solid curve, red dash curve and purple dash-dotted curve respectively. Additionally, we use the solid black line in the figure to show the repeaterless bound~\cite{plob}. 
	
	As shown in Fig.~(\ref{data}a) where the infinite-data case is considered, applying asymmetric signal intensities can always help generate positive key rates for all tested losses. Moreover, at the total loss of 50 dB, the experimental key rate with asymmetric signal intensities is as high as $1.67\times10^{-5}$, even beating the repeaterless bound. However, the key rates of the other two strategies are always lower than the bound. Even worse, no secret keys can be extracted at 56 dB total loss in the adding-loss scenario. If no compensation is applied, there exists positive key rate only when the total loss is 40 dB. In the finite-data case, as shown in Fig.~(\ref{data}b), again, the key rates with asymmetric signal intensities are always higher than the key rates with adding extra losses or applying no compensation. At the total loss of 56 dB, the experimental key rate with asymmetric signal intensities is $3.17\times10^{-7}$ while no keys can be generated with the other two strategies. At 50 dB, the experimental key rate with asymmetric intensities is $6.97\times10^{-6}$, about 30 times of the key rate in the adding-loss scenario.  At 40 dB, the key rate with no compensation is still positive but very small in simulation. However, due to fluctuations in experiment, we could not obtain any keys in the finite-data scenario if no compensation is applied. Note that in Fig.(\ref{data}a), the experiment key rates are always lower than the simulations (except at 40 dB loss). This is due to the fact that the all the intensities are optimized based on finite-data scenario, while the simulations take the intensities optimized for infinite-data scenario.
	
	Overall, the experimental results are consistent with the simulations. As indicated in Fig.(\ref{data}), the distance coverage of TFQKD over optical channels with asymmetric losses is significantly diminished if no compensation is made. Deliberately adding extra losses to compensate the asymmetry could help increase the key rate to some extent, but is not comparable to the strategy of using asymmetric signal intensities. By allowing Alice and Bob to set asymmetric intensities, the secure key rate of TFQKD with asymmetric channel losses can be dramatically increased. The higher the loss, the more key rate enhancement the asymmetric-intensity strategy can achieve. Besides the advantage of providing higher key rate, the asymmetric-intensity strategy is also more convenient and efficient to implement. Especially in a network setting, the adding-loss strategy requires that every user should prepare different compensation losses inside his/her station for different connections. While for the asymmetric-intensity strategy, the users only have to adjust their signal intensities for all different connections. Even when new users join the network, no system modifications are required for the old users. Therefore, a straightforward application of our demonstration in this work can be the study of Sagnac-loop based QKD network.

	In summary, we have demonstrated the first proof-of-principle experiment of TFQKD over optical channels with asymmetric losses. Sagnac interferometer is applied for the auto phase stabilization. Our experiment shows that, compensation strategies are necessary for TFQKD with asymmetric channel losses. Two strategies have been tested, that are applying asymmetric signal intensities or adding extra losses to make the channel loss symmetric again. Compared with the latter strategy, applying asymmetric signal intensities provides much better secure key rate for TFQKD with asymmetric channel losses. It keeps the major advantage of TFQKD, i.e., surpassing the repeaterless bound, and significantly enlarges the distance coverage. Our implementation provides the first experimental study of TFQKD with asymmetric channel losses and shows the feasibility of applying TFQKD to build the long-distance quantum network in reality.  

	We thank Marcos Curty, Feihu Xu and Reem Mandil for their helpful discussion. We thank funding from NSERC, MITACS, CFI, ORF, the Royal Bank of Canada and Huawei Technology Canada Inc.

\end{document}